\documentclass[apl,aip,graphicx]{revtex4-1}

\usepackage{graphicx}% Include figure files
\usepackage{dcolumn}% Align table columns on decimal point
\usepackage{bm}
\usepackage{subfigure}
\usepackage{mathrsfs}
\usepackage{amsmath}
\usepackage{amssymb}
\usepackage[english]{babel}
\draft % marks overfull lines with a black rule on the right
\newcommand{\rmnum}[1]{\romannumeral #1}
\newcommand{\Rmnum}[1]{\expandafter\@slowromancap\romannumeral #1@}

\begin{document}

%\preprint{}

\title{Hall effect in the extremely large magnetoresistance semimetal WTe$_2$} %Title of paper



\author{Yongkang Luo\footnote[1]{Electronic address: ykluo@lanl.gov}}
\affiliation{Los Alamos National Laboratory, Los Alamos, New Mexico 87545, USA.}
\author{H. Li}
\affiliation{Beijing National Laboratory for Condensed Matter Physics, Institute of Physics, Chinese Academy of Science, Beijing 100190, China.}
\author{Y. M. Dai}
\affiliation{Los Alamos National Laboratory, Los Alamos, New Mexico 87545, USA.}
\author{H. Miao}
\author{Y. G. Shi}
\affiliation{Beijing National Laboratory for Condensed Matter Physics, Institute of Physics, Chinese Academy of Science, Beijing 100190, China.}
\author{H. Ding}
\affiliation{Beijing National Laboratory for Condensed Matter Physics, Institute of Physics, Chinese Academy of Science, Beijing 100190, China.}
\affiliation{Collaborative Innovation Center of Quantum Matter, Beijing 100084, China.}
\author{A. J. Taylor}
\author{D. A. Yarotski}
\author{R. P. Prasankumar}
\author{J. D. Thompson}
\affiliation{Los Alamos National Laboratory, Los Alamos, New Mexico 87545, USA.}


\date{\today}

\begin{abstract}
We systematically measured the Hall effect in the extremely large magnetoresistance semimetal WTe$_2$. By carefully fitting the Hall resistivity to a two-band model, the temperature dependencies of the carrier density and mobility for both electron- and hole-type carriers were determined. We observed a sudden increase of the hole density below $\sim$160~K, which is likely associated with the temperature-induced Lifshitz transition reported by a previous photoemission study. In addition, a more pronounced reduction in electron density occurs below 50~K, giving rise to comparable electron and hole densities at low temperature. Our observations indicate a possible electronic structure change below 50~K, which might be the direct driving force of the electron-hole ``compensation'' and the extremely large magnetoresistance as well. Numerical simulations imply that this material is unlikely to be a perfectly compensated system.
\end{abstract}

%71.20.Be Transition metals and alloys
%71.55.Ak Metals, semimetals, and alloys
%72.15.-v Electronic conduction in metals and alloys

\maketitle

The past a few years have witnessed the discovery of a series of new non-magnetic compounds with very large magnetoresistance (MR)\cite{MunE-PtSn4MR,WangK-NbSb2GMR,Ali-WTe2XMR,Ong-Cd3As2,Weng-TmPn}, even much larger than those traditional giant magnetoresistance (GMR) in thin film metals\cite{Egelhoff-GMR} and colossal magnetoresistance (CMR) in Cr-based chalcogenide spinels\cite{Ramirez-CrCMR} and Mn-based perovskites\cite{Jin-MnCMR}. With many potential applications in magnetic field sensors, read heads, random access memories, and galvanic isolators\cite{Daughton-GMRAppl}, these newly discovered extremely large magnetoresistance (XMR) materials have attracted enormous interest. Several mechanisms have been proposed as the origin of the XMR in these materials, {\it e.g.}, (\rmnum{1}) topological protection from backward scattering mechanism, and (\rmnum{2}) electron-hole compensation mechanism. The former has been realized in Dirac semimetals like Cd$_3$As$_2$\cite{Ong-Cd3As2,WangJ-Cd3As2SdH} and Weyl semimetals such as $TmPn$ (where $Tm$ = Ta or Nb, and $Pn$ = As or P)\cite{Weng-TmPn,Lv-TaAsPRX,Shekhar-NbP,Nirmal-NbAs,LuoY-NbAsSdH,Shekhar-TaPLMR}, while the latter stems from a multi-band effect\cite{Singleton-Band}: although no net current flows in the $\textbf{y}$-direction, currents in the $\textbf{y}$-direction carried by a particular type of carrier may be non-zero. These transverse currents experience a Lorentz force that is antiparallel to the $\textbf{x}$-direction. This backflow of carriers provides an important source of magnetoresistance, which is most pronounced in semimetals like Bi\cite{Alers-BiMR} and graphite\cite{Du-BiCMIT} where electrons and holes are compensated.

Recently, the observation of XMR in high-purity WTe$_2$ has triggered great enthusiasm\cite{Ali-WTe2XMR}. Angle-resolved photoemission spectroscopy (ARPES) studies have revealed the coexistence of multiple electron and hole Fermi surfaces (FSs) with the total size of electron pockets close to that of hole pockets\cite{Pletikosi-WTe2ARPES,JiangJ-WTe2ARPES,WuY-WTe2ARPES}, which is further supported by Shubnikov-de Hass (SdH) quantum oscillation\cite{ZhuZW-WTe2SdH,CaiP-WTe2PreSdH} measurements, reminiscent of the electron-hole compensation mechanism for WTe$_2$.
Furthermore, ARPES measurements also manifested a temperature-induced Lifshitz transition at about 160 K, above which all the hole bands sink below the Fermi level\cite{WuY-WTe2ARPES}. This raises the possibility that the temperature-induced Lifshitz transition may be the driving force for the electron-hole compensation as well as the XMR at low temperatures. On the other hand, a more recent transport study reported that the mass anisotropy grows sharply below $\sim$50~K\cite{Thoutam-WTe2Aniso}, closely following the temperature dependence of the MR. This mass anisotropy enhancement has been associated with a change in the electronic structure that is believed to play a key role in turning on the XMR in WTe$_2$. Direct evidence for these views may be produced by a thorough investigation into the temperature dependence of the carrier density that can be determined by Hall effect measurements.

In this {\it Letter}, we performed systematic measurements of electrical transport properties ($\rho_{xx}$ and $\rho_{yx}$) on high-quality WTe$_2$ single crystals. A careful fitting of $\rho_{yx}(B)$ to a two-band model yields the temperature dependencies of the carrier density and mobility for both electron- and hole-type of carriers. The signature of the temperature-induced Lifshitz transition at $\sim$160~K was observed. In addition, the electron carrier density significantly drops below 50~K, leading to a \emph{nearly} compensated situation at low temperature which is probably the cause of the XMR. These experiments pose an interesting ``paradox'' between a quadratic-law for $\rho_{xx}(B)$ and a non-linear $\rho_{yx}(B)$. Further numerical simulations suggest that WTe$_2$ is unlikely to be a perfectly compensated system. Our results reveal the mechanism of the XMR in WTe$_2$, shedding light on this peculiar material with great potential for electronic device applications.

\begin{figure}
\includegraphics[width=8.5cm]{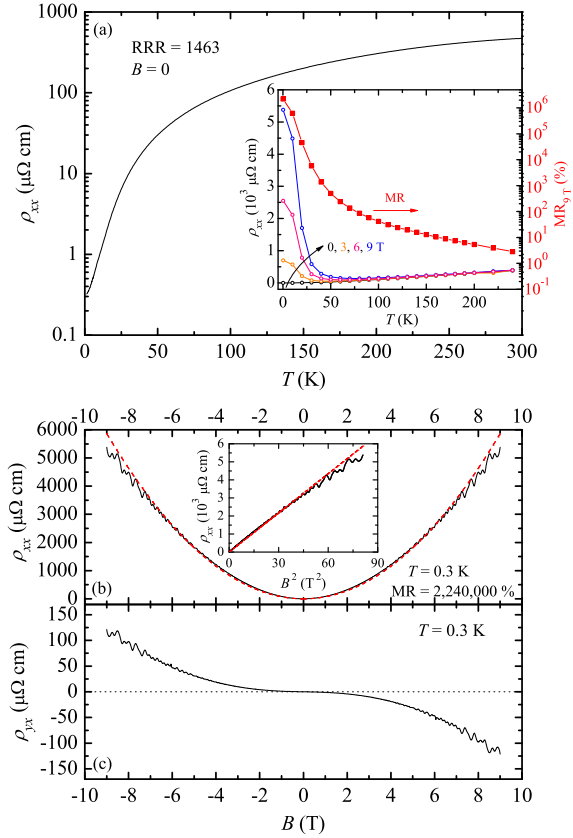}
\caption{\label{Fig.1} (a) Temperature dependence of the resistivity $\rho_{xx}(T)$ of WTe$_2$ in the absence of a magnetic field. The inset shows a comparison of $\rho_{xx}(T)$ at $B$ = 0, 3, 6 and 9 T (left axis), and \%MR$_{9 T}$$\equiv$100$\times$$[\rho_{xx}(9~\text{T})$$-$$\rho_{xx}(0)]/\rho_{xx}(0)$ (right axis). (b) and (c) display the field dependencies of $\rho_{xx}$ and $\rho_{yx}$ at 0.3 K, respectively. The inset of (b) shows $\rho_{xx}(B)$ plotted in the $B^2$ scale with the red-dashed line as a linear guide to the eyes. }
\end{figure}

The experimental details and sample characterization have been addressed in \textbf{Supplemental Material} (\textbf{SM})\cite{SM}. The main panel of Fig.~\ref{Fig.1}(a) shows the temperature dependent resistivity of WTe$_2$ measured without an external magnetic field. Our sample is characterized by a huge residual resistance ratio [RRR$\equiv$$R(300~\text{K})/R(1.3~\text{K})$=1463], which is among the highest reported\cite{Ali-WTe2XMR,CaiP-WTe2PreSdH,WangJ-WTe2Angel,ZhuZW-WTe2SdH,WuY-WTe2ARPES}. In Figs.~\ref{Fig.1}(b) and \ref{Fig.1}(c), we respectively present $\rho_{xx}$ and $\rho_{yx}$ at $T$ = 0.3 K as a function of magnetic field $B$. The field dependence of $\rho_{xx}$ roughly follows a quadratic behavior [red-dashed line in Fig.~\ref{Fig.1}(b)] without any trend of saturation\cite{Ali-WTe2XMR}. The resulting extremely large magnetoresistance \%MR$\equiv$100$\times$$[\rho_{xx}(B)$$-$$\rho_{xx}(0)]/\rho_{xx}(0)$ reaches 2,240,000\% at 9~T, comparable with the values in previous reports\cite{Ali-WTe2XMR,CaiP-WTe2PreSdH,WangJ-WTe2Angel,ZhuZW-WTe2SdH,WuY-WTe2ARPES}. The large magnitudes of RRR and MR guarantee the high quality of our WTe$_2$ single crystal. Another important feature of the magneto-transport properties in WTe$_2$ is the large SdH quantum oscillations, observed in both $\rho_{xx}(B)$ [Fig.~\ref{Fig.1}(b)] and $\rho_{yx}(B)$ [Fig.~\ref{Fig.1}(c)]. By taking the Fast Fourier Transform (FFT) of the oscillatory part $\Delta\rho_{xx}$=$\rho_{xx}$$-$$\langle\rho_{xx}\rangle$ (where $\langle\rho_{xx}\rangle$ is the non-oscillatory background), we derived the extremal cross-sectional areas $S_F$ of the FSs that are similar to Zhu {\it et al.} \cite{ZhuZW-WTe2SdH}. \textbf{SM}\cite{SM} provides more details about SdH oscillations.

\begin{figure}
\includegraphics[width=16cm]{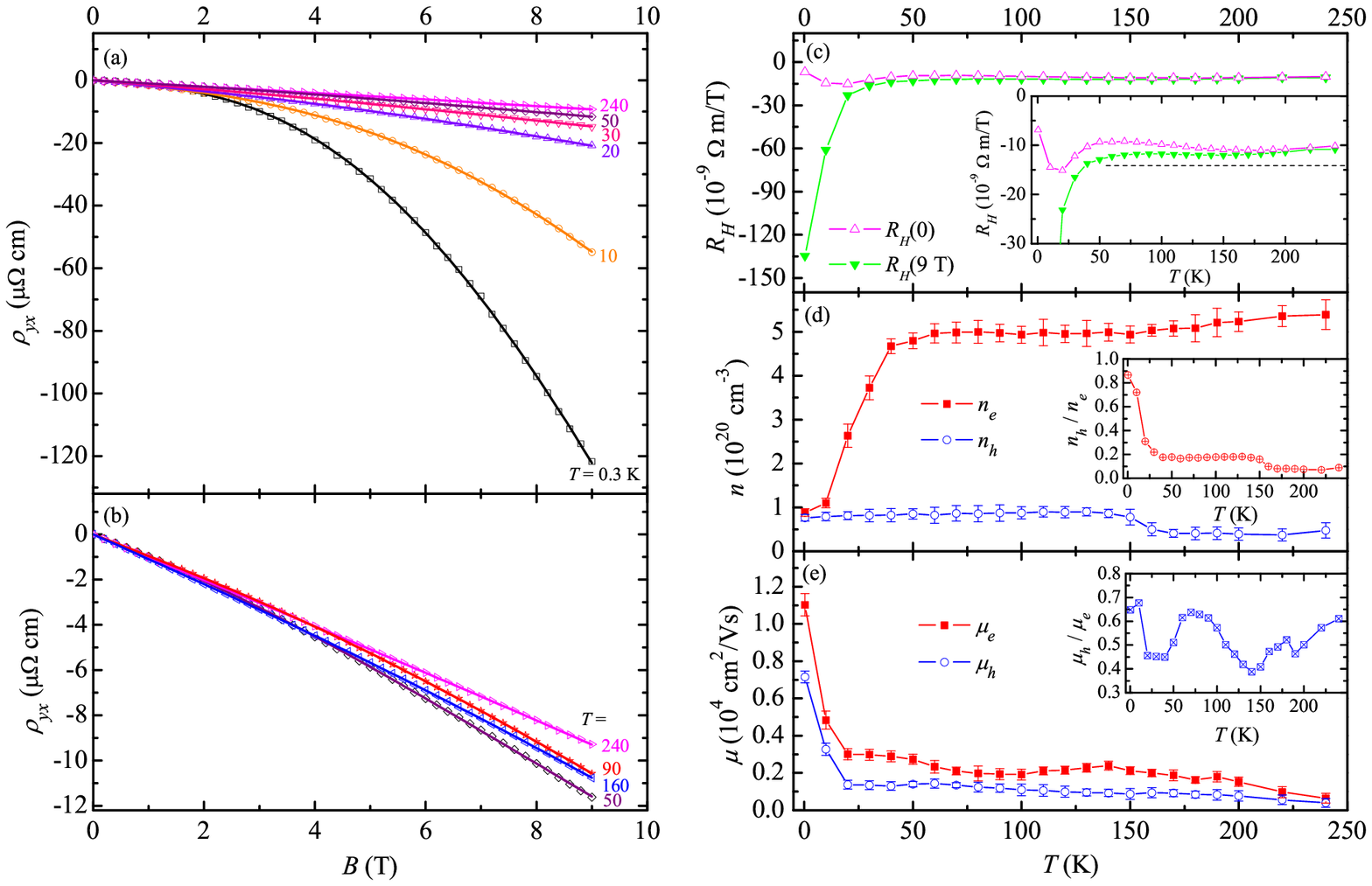}
\caption{\label{Fig.2} (a) and (b) show field dependencies of the Hall resistivity at various temperatures. The curve for 0.3~K is the non-oscillating background $\langle\rho_{yx}\rangle$. The open symbols are experimental data, while the solid lines running through the data are numerical fitting to a two-band model Eq.~(\ref{Eq.2b}). (c) Hall coefficient $R_H$ as a function of $T$. The solid symbols signify the $R_H$ defined by $\rho_{yx}/B$ at $B$ = 9~T, while the open symbols stand for $R_H$ determined from the initial slope of $\rho_{yx}(B)$ at $B\rightarrow$ 0. The dashed line in the inset is a guide line to eyes. (d) and (e) display the temperature dependent carrier density and mobility, respectively. The insets to (d) and (e) show the plot of $n_h/n_e$ and $\mu_h/\mu_e$ versus $T$.}
\end{figure}

Figures~\ref{Fig.2}(a) and \ref{Fig.2}(b) depict the field dependence of $\rho_{yx}$ at selected temperatures ranging from 0.3 K to 240 K. It is evident that for all temperatures, $\rho_{yx}$ is negative, manifesting an electron-dominant characteristic. We also note that $\rho_{yx}$ is not a linear function of $B$. This non-linearity of $\rho_{yx}(B)$ becomes more pronounced at low temperature, and is also reflected in the temperature dependent Hall coefficient $R_H$ as shown in Fig.~\ref{Fig.2}(c). The solid symbols in Fig.~\ref{Fig.2}(c) signify $R_H$(9~T) defined by $\rho_{yx}/B$ at $B$ = 9~T, and the open symbols stand for $R_H(0)$ determined from the initial slope of $\rho_{yx}(B)$ near $B$ = 0. At high temperatures, the two curves more or less overlap, whereas they apparently diverge below $\sim$160~K. This divergence is more significant below 20~K where $R_H$(9~T) dramatically decreases but $R_H(0)$, on the contrary, starts to increase. All these arise from the multi-band effect in this material. In fact, according to the recent band structure calculations\cite{Ali-WTe2XMR}, ARPES\cite{WuY-WTe2ARPES} and SdH\cite{ZhuZW-WTe2SdH} results, the FS of WTe$_2$ at low temperatures consists of two electron-like sheets and two hole-like sheets on each side of the $\boldsymbol{\Gamma}$-$\textbf{X}$ line in the 1st Brillouin zone.

For simplicity, here we adopt a two-band model with one electron-and one hole-bands. 
%This model has been successfully used to describe the Hall effect in multi-band iron-pnictides\cite{Albenque2009,Albenque2010}, and we will see that it is sufficient to explain the Hall effect in the full temperature range. 
The total conductivity tensor $\boldsymbol{\sigma}$ is conveniently expressed in the complex representation\cite{Ali-WTe2XMR},
\begin{equation}
\boldsymbol{\sigma}=e\left [\frac{n_e\mu_e}{1+i\mu_eB}+\frac{n_h\mu_h}{1-i\mu_hB}\right ],
\label{Eq.1}
\end{equation}
where $n$ and $\mu$ are carrier density and mobility, and the subscript $e$ (or $h$) denotes electron (or hole). Converting this equation into the resistivity tensor ($\boldsymbol{\rho}$ = $\boldsymbol{\sigma}^{-1}$), $\rho_{xx}$ and $\rho_{xy}$(=$-\rho_{yx})$ are the real and imaginary parts of $\boldsymbol{\rho}$, respectively, i.e.,
\begin{subequations}
\label{Eq.2}
\begin{align}
\rho_{xx}(B)&=\text{Re}(\boldsymbol{\rho})=\frac{1}{e}\frac{(n_h\mu_h+n_e\mu_e)+(n_h\mu_e+n_e\mu_h)\mu_h\mu_eB^2}{(n_h\mu_h+n_e\mu_e)^2+(n_h-n_e)^2
\mu_h^2\mu_e^2B^2}, \label{Eq.2a}\\
\rho_{yx}(B)&=-\text{Im}(\boldsymbol{\rho})=\frac{B}{e}\frac{(n_h\mu_h^2-n_e\mu_e^2)+(n_h-n_e)\mu_h^2\mu_e^2B^2}{(n_h\mu_h+n_e\mu_e)^2+(n_h-n_e)^2
\mu_h^2\mu_e^2B^2}. \label{Eq.2b}
\end{align}
\end{subequations}
All the $\rho_{yx}(B)$ data measured at different temperatures can be well fitted to Eq.~(\ref{Eq.2b}), with $n_e$, $n_h$, $\mu_e$ and $\mu_h$ being fitting parameters. Some representative results are displayed in Figs.~\ref{Fig.2}(a) and \ref{Fig.2}(b). The fitting derives the temperature dependencies of carrier density and mobility for both electron- and hole-type carriers, shown in Figs.~\ref{Fig.2}(d) and \ref{Fig.2}(e), respectively. An important feature of $n_h(T)$ is that it increases abruptly below $\sim$160 K. At 170 K, $n_h=0.41\times 10^{19}$ cm$^{-3}$. It increases to 0.86$\times 10^{19}$ cm$^{-3}$ at 140 K, more than twice of the value at 170 K. Temperature dependent ARPES measurements have pointed out that a temperature-induced Lifshitz transition is likely to occur at about 160~K, below which the hole pockets appear\cite{WuY-WTe2ARPES}. The observed upturn of $R_H(T)$ [see inset to Fig.~\ref{Fig.2}(c)] and the enhancement of $n_h(T)$ below this critical temperature are consistent with the temperature-induced Lifshitz transition. Furthermore, we noticed that $n_e(T)$ decreases drastically below 50 K. More interestingly, $n_e(T)$ and $n_h(T)$ tend to be comparable at low temperature. This is better seen in the ratio $n_h/n_e$ shown in the inset to Fig.~\ref{Fig.2}(d). In particular, at 0.3 K, $n_e$ = 8.82 $\times10^{19}$~cm$^{-3}$ and $n_h$ = 7.64 $\times10^{19}$ cm$^{-3}$. These values are very close to the carrier densities determined from SdH measurements\cite{ZhuZW-WTe2SdH} (See also in \textbf{SM}). Furthermore, $T$ = 50~K is also the characteristic temperature below which the magnetoresistance starts to increase dramatically, see the inset of Fig.~\ref{Fig.1}(b). Thermopower measurements also exhibit a sign change from being positive to negative upon cooling through 50~K\cite{Kabashima-WTe2Trans,WuY-WTe2ARPES}. Our recent ultrafast optical pump-probe spectroscopic measurements also revealed that the timescale governing electron-hole recombination, which is sensitive to the electronic structure, shows a strong anomaly at $\sim$50~K\cite{DaiYM-WTe2UFOS}. All these experimental results suggest a potential electronic structure change below 50~K, which is likely the direct driving mechanism of the electron-hole ``compensation" and the XMR at low temperature as well. Since no systematic temperature dependent ARPES measurements have been done for temperatures below 50 K, the nature of this electronic structure change remains an open question that needs to be clarified in the future.

\begin{figure}
\includegraphics[width=8.5cm]{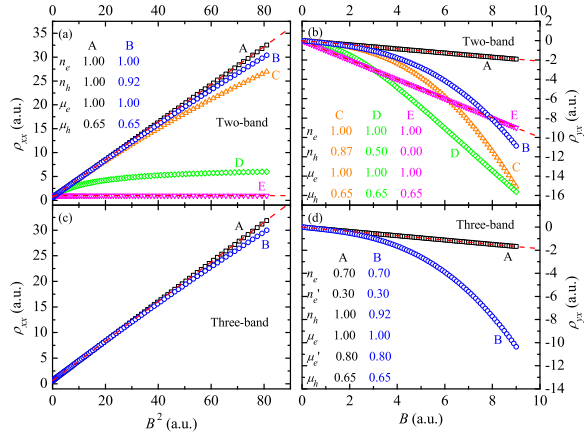}
\caption{\label{Fig.3} Numerical simulations of $\rho_{xx}(B)$ (left panels) and $\rho_{yx}(B)$ (right panels). The parameters are listed in panels (a), (b) and (d). The red dashed lines are the linear guides to the eyes. (a) and (b) are calculated with a two-band model. (c) and (d) are calculated with a three-band model.}
\end{figure}

Finally, we realized an interesting ``paradox" between the behaviors of $\rho_{xx}(B)$ and $\rho_{yx}(B)$. According to Eq.~(\ref{Eq.2a}), the condition for $\rho_{xx}(B)$ to increase as $B^2$ is $n_e$ = $n_h$, which is the case that electrons and holes are \emph{perfectly} compensated by each other\cite{Ali-WTe2XMR}. However, this inevitably leads to linearity of $\rho_{yx}(B)$ [see Eq.~(\ref{Eq.2b})], which is apparently at odds with the experimental data shown in Fig.~\ref{Fig.2}. This paradox can not be reconciled by tuning the carrier mobilities or introducing additional bands, but can only be resolved by slightly unbalancing $n_h$ and $n_e$.

To better clarify this, we performed numerical simulations as displayed in Fig.~\ref{Fig.3}. All quantities depicted here are in arbitrary units. In Figs.~\ref{Fig.3}(a) and \ref{Fig.3}(b), we calculated $\rho_{xx}(B)$ and $\rho_{yx}(B)$ based on the two-band model Eq.~(\ref{Eq.1}). The black curves A signify the perfect compensation condition, with the parameters $n_e$ = 1.00, $n_h$ = 1.00, $\mu_e$ = 1.00, and $\mu_h$ = 0.65, and indeed, we obtained quadratic $\rho_{xx}(B)$ and linear $\rho_{yx}(B)$. Slightly decreasing $n_h$, {\it e.g.} 0.92, we found that $\rho_{xx}(B)$ weakly deviates from the quadratic-law and meanwhile $\rho_{yx}(B)$ is strongly non-linear (see curve B), which is qualitatively similar to the experimental data shown in Figs.~\ref{Fig.1}(b) and \ref{Fig.1}(c). As a comparison, we also show in Figs.~\ref{Fig.3}(c) and \ref{Fig.3}(d) the simulations based on a three-band model. The total conductivity tensor now becomes
\begin{equation}
\boldsymbol{\sigma}=e\left [\frac{n_e\mu_e}{1+i\mu_eB}+\frac{n_e'\mu_e'}{1+i\mu_e'B}+\frac{n_h\mu_h}{1-i\mu_hB}\right ],
\label{Eq.3}
\end{equation}
and the condition for a perfect electron-hole compensation changes to $n_e$$+$$n_e'$=$n_h$. Nevertheless, the compensation condition still leads to the aforementioned paradox. To note, the situation can not be improved by introducing a fourth band or even more bands (data not shown). All these suggest that WTe$_2$ is not a purely compensated system, whereas we should admit that a slight mismatch between $n_e$ and $n_h$ does not strongly affect the XMR [Curves B and C in Fig.~\ref{Fig.3}(a)]. XMR disappears when $n_e$ and $n_h$ are severely unbalanced, seeing the curves D and E in Fig.~\ref{Fig.3}(a).

To summarize, we analysed the Hall effect in the extremely large magnetoresistance semimetal WTe$_2$ within a two-band model, and derived the temperature dependencies of the carrier density and mobility for both electron- and hole-type carriers. Below $\sim$160~K, the hole carrier density abruptly increases, consistent with a temperature-induced Lifshitz transition observed by a previous ARPES study. Moreover, the electron-type carrier density decreases sharply below 50~K, and at low temperature, the carrier densities of electrons and holes become comparable. Our results indicate a possible electronic structure change at about 50~K, which is likely to drive the electron-hole ``compensation" that promotes the extremely large magnetoresistance. We also performed numerical simulations of $\rho_{xx}(B)$ and $\rho_{yx}(B)$ based on multi-band models, and our calculations suggest that this material is unlikely to be a perfectly compensated system.

We thank John Bowlan, Pamela Bowlan, Brian McFarland, and F. Ronning for insightful discussions. Work at Los Alamos was performed under the auspices of the U.S. Department of Energy, Division of Materials Science and Engineering, and Center for Integrated Nanotechnologies. Work at IOP CAS was supported by the Strategic Priority Research Program (B) of the Chinese Academy of Sciences (Grant No. XDB07020100) and the National Natural Science Foundation of China (No. 11274367 and 11474330). %Y. Luo acknowledges a Director's Postdoctoral Fellowship supported through the LANL LDRD program.

Y. Luo and H. Li contributed equally to this work.

%\bibliography{biblio}


\end{document}